# A Hybrid System based on Multi-Agent System in the Data Preprocessing Stage

Kobkul Kularbphettong, Gareth Clayton and Phayung Meesad
The Information Technology Faculty,
King Mongkut's University of Technology
North Bangkok, Thailand

*Abstract*—We describe the usage of the Multi-agent system in the data preprocessing stage of an on-going project, called e-Wedding. The aim of this project is to utilize MAS and various approaches, like Web services, Ontology, and Data mining techniques, in e-Business that want to improve responsiveness and efficiency of systems so as to extract customer behavior model on Wedding Businesses. However, in this paper, we propose and implement the multi-agent-system, based on JADE, to only cope data preprocessing stage specified on handle with missing value techniques. JADE is quite easy to learn and use. Moreover, it supports many agent approaches such as agent communication, protocol, behavior and ontology. This framework has been experimented and evaluated in the realization of a simple, but realistic. The results, though still preliminary, are quite.

*Keywords-component; multi-agent system, data preprocessing stage, Web services, ontology, data mining techniques, e-Wedding, JADE.*

## I. INTRODUCTION

Data Preprocessing is one of the significant factors that affects on the success of Data Mining and Machine Learning approaches. Generally, data preprocessing stage represents the quality of data. The more incorrect and incomplete data presents, the more result is unreliable. Moreover, data preprocessing task is time consuming because it includes many phases like data cleaning, data integrating, data transforming and data reducing. However, the best performance of the data pre-processing algorithms is relied on the nature of each data set. Hence, it would be nice if it have the interested methodology to adapt for choosing the best performance of the data preprocessing algorithms for each data set.

Recently, although there are much of researches applied MAS (Multi-agent system) in a wide range of problem in Data Mining and Machine Learning techniques, very few researches are focused on using MAS in the data preprocessing step. A multi agent system is a computational system, or a loosely coupled network in which two or more agents interact or work together to perform a set of tasks or to satisfy a set of goals. Each agent is considered as a locus of a problem-solving activity which operates asynchronously with respect to the other agents [1]. Therefore, in this paper we propose the MAS framework of an on-going project, called e-Wedding, that merely focuses on using MAS to handle in the problems of data preprocessing stage.

The remainder of this paper is organized as follows. Section 2 reviews about related literatures and research works in the use of multi-agent system for data mining. Section 3 presents the related methodologies used in this work. Section 4 presents the experimental results based on the purposed model based on multi-agent framework. This prototype demonstrates how to success for adapting multi-agent system in data preprocessing stage. Finally, we conclude the paper with future research issues in section 5.

## II. RELATED WORKS

A literature search shows that most of the related researches have been deployed multi-agent to cope with the data mining techniques specified on the data preprocessing algorithms by following this:

According to [5], they showed a prototype of the system using the JADE platform in the context of travel industry. Furthermore, other research works show that agent technologies are deployed as a significant tool for developing e-Commerce applications [2]-[7]. Hence, multi-agent technology, a promising approach, trends to handle internet transaction for customers.

Moreover, other researchers propose an agent-based framework representing in various ways that are related with data mining techniques [8]-[10]. For instances, Chien-Ho Wu, Yuehjen E. Shao, Jeng-Fu Liu, and Tsair-Yuan Chang[11] applied agent technology to collect and integrate data distributed over various computing platforms to facilitate statistical data analysis in replacing the missing values by using either Approximate Bayesian Bootstrap (ABB) or Ratio Imputation and using MAS improves the execution time at different by focusing on spatial knowledge in order to extract knowledge in Predictive Modeling Markup Language (PMML) format [9].

From previous literature works, it appears that there are many research studies exploiting various techniques blended with multi-agent technology and data mining techniques. Consequently, in order to success on e-Commerce, agent should have abilities to perform as a behalf of user to handle with business tasks such as planning, reasoning and learning. Also, data mining techniques is the important way to make a reason for agent under uncertainty and with incomplete information situations. Notwithstanding, data preprocessing step acts as an crucial task to filter and select suitable information before processing any mining algorithms.

### III. THE METHODOLOGIES

In this section, we illustrate the specified methodologies used in this project but it is only focused on the approaches using in the data preprocessing stage specified in dealing with missing values.

#### A. Multi-Agent System

Agent is the software program that enables to autonomous action in some environment so as to meet its design objectives. According to N. R. Jennings and M. Wooldridge [12], the essential characters of each agent are following: reactive, pro-active, autonomous, object-oriented and social ability. Each agent can play as a behalf of the user and execute the particular task. Also, Padghan and Winikopff [13] described that the concept of Agent refers to an entity acting on behalf of other entities or organizations and having the ability to perceive relevant information, and following and handling the objectives to be accomplished. However, in open and dynamic environment like internet, a multi-agent system is one of the promising means to help reduce cost, increase efficiency, reduce errors and achieve optimal deal.

There are two issues related to the design of MAS: Agent Communication Language and agent development platform. The former concerns with the message interchange between different agent such as KQML, and FIPA ACL. The latter is related with the platform development to provide an effective framework, such as IBM Aglets, ObjectSpace Voyager and etc, for the dispatching, communications, and management of multiple agents in the open and dynamic environment.

For this proposed project, JADE (Java Agent Development Framework) will be deployed as the prototype development tool. JADE (Java Agent Development Framework) is a software environment fully implemented in JAVA language aiming at the development of multi-agent systems that comply with FIPA specifications [14]. The goal of JADE is to simplify development while ensuring standard compliance through a comprehensive set of system services and agents. Each running instance of the JADE runtime environment is called a container as it can contain several agents. The set of active containers is called a platform. A single special container must always be active in a platform and all other containers register with it as soon as they start. Hence, the development framework based on JADE is considered very suitable for implementing applications that require distributing computation tasks over the network.

#### B. Data Preprocessing Techniuques

Data pre-processing is an often neglected but important step in the data mining process, as depicted in figure. 1. The phrase "Garbage In, Garbage Out" is particularly applicable to data mining and machine learning projects [15].

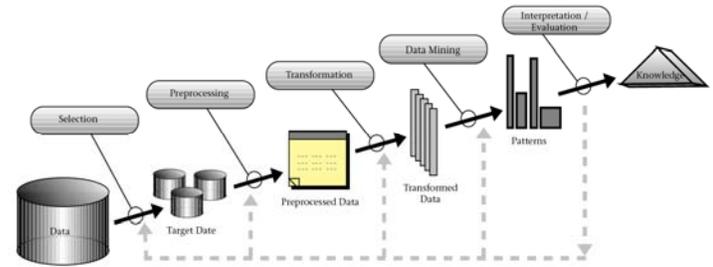

Figure 1. The data mining steps [16].

There are many tasks in data preprocessing like data cleaning, data transformation, data reduction, data integration and etc. Nevertheless, in this paper, we emphasizes on the data cleaning stage so as to handle missing data.

Missing data is a common problem for data quality in real datasets. However, there are several methods for handling missing data and according Little and Rubin[17], missing data treatment can be divided to three categories a) Ignoring and discarding data, which known discarding method can be categorized to the complete case analysis method and the pair wise deletion method. The former discards all instances with missing data and the latter discards only the instances with high level of missing data, determined the extent of missing data before b) Parameter estimation, which Maximum likelihood procedures that use variants of the Expectation-Maximization algorithm can cope estimation parameter in the presence of missing and c) Imputation techniques, which missing values are filled with estimated ones based on information available in the data set.

Also, there are some popular missing data treatment methods that researcher have to choose by following this:

- *Mean or mode substitution*: replacing all missing data with the mean (numeric attribute) or mode (nominal attribute) of all observed cases. However, the drawbacks of using these methods are to changing the characteristic of the original dataset and ignoring the relationship among attributes that affect on the performance of the data mining algorithms.

- *Regression substitution*: replacing all missing data with a statistics relied on the assumption of linear relationship between attributes.

- *Hot deck imputation:* replacing all missing data with an estimated distribution from the current data. In random hot deck, a missing value is replaced by a observed value (the donor) of the attribute chosen randomly, similarly to hot deck, but in cold deck methods, the imputed value must be different from the current data source.

- *KNN Imputation*: replacing all missing data with k-nearest neighbor algorithm that determines the

similarity of two instances by using a distance function.
- Classification methods: replacing all missing data with classification models, like decision tree, C4.5 and etc, and using all relevant features as predictors.

## IV. THE PURPOSED FRAMEWORK AND EXPERIMENTAL RESULTS

This section displayed the purposed framework of this project and compares the result of the chosen missing value algorithms. For illustration of framework as figure 2-4 [18, 19], we select the wedding businesses and its environment. There are several issues in this system such as multi-agent system, web services, ontology, and data mining techniques, as shown in figure 3, but in this paper we present merely a multi-agent system dealing with data preprocessing steps and focusing on the missing value techniques.

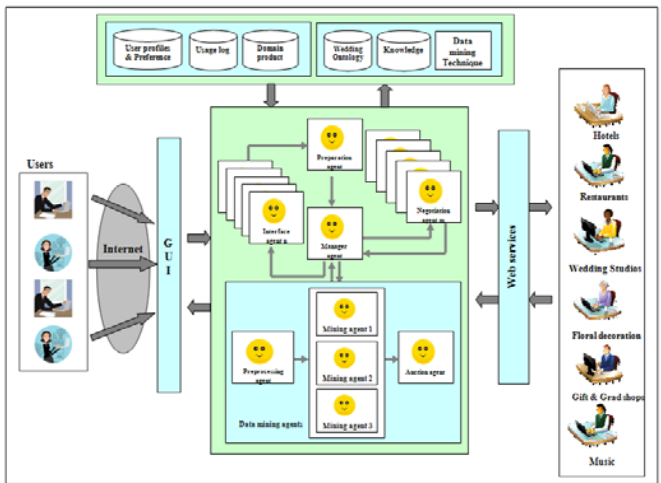
Figure 2. The purposed architecture of the e-Wedding system.

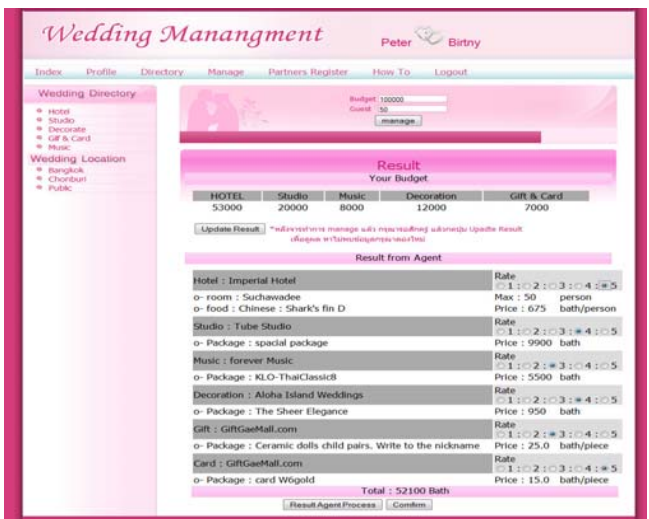
Figure 3. The web page in e-Wedding System.

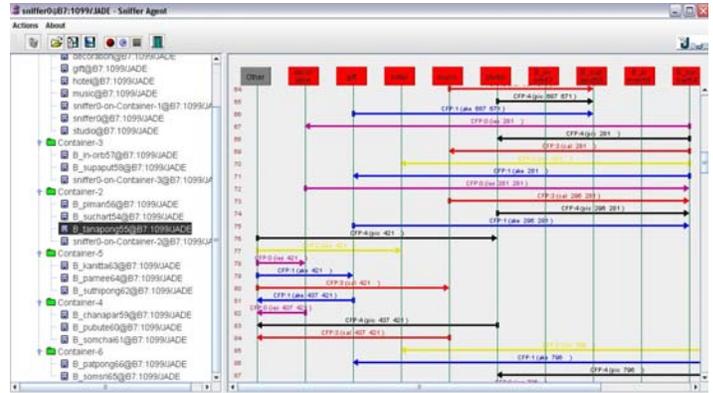
Figure 4. The operation of JADE in e-Wedding System.

Recently, there are a lot of algorithms dealing with missing data. The basic approaches about these popular algorithms have been introduced above section. In this paper, we implement a composite imputation method between hot deck and nearest neighbor methods based on mean substitution, shown figure 4.

Hot deck imputation technique is commonly used in statistic for item non response. The main concept of the hot deck method is to use the current data, called donors, to provide imputed values for records with missing values. The procedure through which we find the donor that matches the record with missing values is different according to the particular techniques used [20].

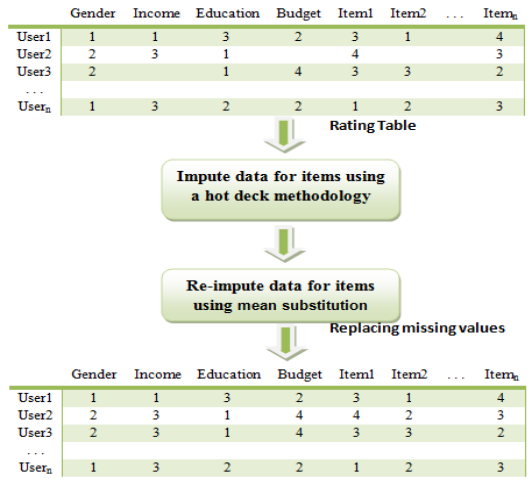
Figure 5. The process of the purposed Imputation technique.

A hot deck imputation method can be described by two factors: the way in which donors are selected for each missing instance and the way in which the weight of the donor is defined for each missing instance [21].

According to Jae Kwang Kim, the first way can determine by the distribution of d equals

$$d = d_{ij}; \quad i \in A_R, j \in A_m \tag{1}$$

where  $A_R$ = the set of indices of the sample respondents
$A_m$ = the set of indices of the sample nonrespondents
And $d_{ij}$ = the number of times that $Y_i$ is used as donor for $Y_j$

The second way can determine the weight of the donor specified for each missing item. For missing item j:

$$Y_{ij} = \sum_{i \in A_R} d_{ij} w_{ij}^* y_i \qquad (2)$$

Let  $w_{ij}^*$ = the fraction of the original weight assigned to donor i as a donor for element j.

Also, to tailor the hot deck imputation process, predictive mean matching is applied to this process. Predicted means are then calculated for both records where the item is missing and records where it is non-missing. Donors for those records requiring imputation are selected by matching on the predicted means, according to some specified distance metric. The imputed value is then the value of the item on the donor record.

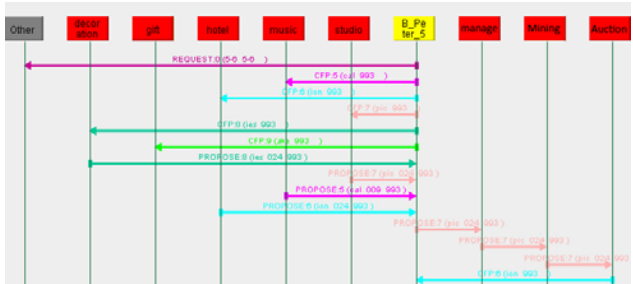

Figure 6. The operation of JADE in the purposed Imputation technique.

## V. CONCLUSION AND FUTURE WORKS

In this paper we presented our preliminary ideas of building multi-agent system with data preprocessing steps by specified in a missing value step, based on e-Wedding system. In the part of MAS, we have implemented this prototype by using JADE platform. JADE is quite easy to learn and use. Moreover, it supports many agent approaches such as agent communication, protocol, behavior and ontology. As for the future work, we need to explore more reasonable and data mining technologies

## REFERENCES

[1] Sandholm, T. and Lesser, "*Advantages of a Leveled Commitment Contracting Protocol.*" Thirteenth National Conference on Artificial Intelligence (AAAI-96), pp. 126--133, Portland, OR.

[2] Bala M. Balachandran and Majigsuren Enkhsaikhan**,** "*Developing Multi-agent E-Commerce Applications with JADE*", Lecture Notes in Computer Science, Springer Berlin / Heidelberg 2009.

[3] Mu-Kun Cao, Yu-Qiang Feng, Chun-Yan Wang, "*Designing Intelligent Agent for e-Business Oriented Multi-agent Automated Negotiation.*" Proceedings of the Fourth International Conference on Machine Learning and Cybernetics, Guangzhou, 18-21 August 2005.

[4] Patrick C. K. Hung, Ji-Ye Mao. "*Modeling of E-negotiation Activities with Petri Nets.*", HICSS 2002.

[5] Y. Yuan, J. B. Rose, N. Archer, and H. Suarga, *"A Web-Based Negotiation Support System."* International Journal of Electronic Markets,1998.

[6] Giacomo Piccinelli, Claudio Bartolini and Chris Preist. *"E-service composition: supporting dynamic definition of process-oriented negotiation"* In Proc. 12th International Workshop on Database and Expert Systems Applications (DEXA 2001), Munich, Germany, September 2001. IEEE Computer Society 2001.

[7] Akkermans, H. "Intelligent E-Business - From Technology to Value.", IEEE Intelligent Systems, 16(4):8-10, 2001.

[8] Huang Xin Li; Chosler, R. "Application of Multilayered Multi-Agent Data Mining Architecture to Bank Domain", Wireless Communications, NetworkingandMobileComputing,2007.InternationalConferenceonVolume, Issue , 21-25 Sept. 2007 Page(s):6721 – 6724.

[9] H. Baazaoui Zghal, S. Faiz, and H. Ben Ghezala, "A Framework for Data Mining Based Multi-Agent: An Application to Spatial Data", World Academy of Science, Engineering and Technology 5 2005.

[10] Zili Zhang, Chengqi Zhang and Shichaozhang, "An Agent-Based Hybrid Framework for Database Mining", Applied Artificial Intelligence, 17:383–398, 2003.

[11] Chien-Ho Wu, Yuehjen E. Shao, Jeng-Fu Liu, and Tsair-Yuan Chang , "On Supporting Cross-Platform Statistical Data Analysis Using JADE", Book Series Studies in Computational Intelligence, Springer Berlin Heidelberg, *issn1860-949X (Print) 1860-9503 (Online)*, Volume 214/2009.

[12] N. R. Jennings and M. Wooldridge. "Software Agents", *IEE Review* 42(1), pages 17-21. January 1996.

[13] Padghan, L. and Winikopff, M., "Developing Intelligent AgentSystems.", Wiley.2004**.**

[14] JADE, Java Agent Development Environment,2006, http://jade.tilab.com

[15] http://en.wikipedia.org/wiki/Data_Pre-processing

[16] http://www.infovis-wiki.net/index.php?title=Image:Fayyad96kdd-process.png

[17] Little, R. J. A. and Rubin, D. B.,"Statistical Analysis with Missing Data" 2nd Edition, John Wiley & Sons, New York, 2002

[18] Kobkul Kilarbphettong, "*e-Negotiation based on Multi-agent ssyetm*",JCSSE 2007 –The international Joint Conference on Computer Science and software Engineer,Thailand.

[19] Kobkul Kilarbphettong**,** Gareth Clayton, and Phayung Meesad**,** "*e-Wedding System based on Multi-System*", Advances in Intelligent and Soft-Computing ,series of Springer, 2010

[20] Andrea Piesse, David Judkins, and Zizhong Fan., "*Item Imputation Made Easy*", Proceedings of the Survey Research Methods Section,2005.

[21] Jae Kwang Kim, "Variance Estimation for Nearest Neighbor Imputation with Application to Census Long Form Data", Proceedings of the Survey Research Methods Section,2002.

### AUTHORS PROFILE

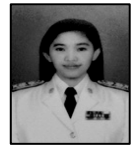

**Kobkul Kularbphettong** received the B.S. degree in Computer Business, M.S. degree in Computer Science. She is Currently Ph.D. Student in Information Technology. Her current research interests Multi-agent System, Web Services, Semantic Web Services, Ontology and Data mining techniques.

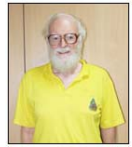

**Dr.Gareth Clayton** is a Statistician, so for IT students any project involving statistics, including the following, but not excluding other areas: Data Mining, Simulation studies, Design of Experiments, Model Fitting Parameter Estimation.

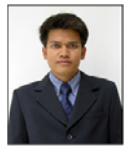

**Phayung Meesad** received the B.S.,M.S., and Ph.D. degree in Electrical Engineering. His current research interests Fuzzy Systems and Neural Networks, Evolutionary Computation and Discrete Control System.